**RESEARCH ARTICLE**

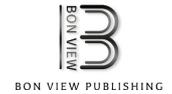

# Prevention of Shoulder-Surfing Attack Using Shifting Condition with the Digraph Substitution Rules

Amanul Islam[1,*] , Fazidah Othman[1], Nazmus Sakib[2] and Hafiz Md. Hasan Babu[3]

[1]Department of Computer Science and Information Technology, University of Malaya, Malaysia
[2]Department of Computer Science and Engineering, Dhaka International University, Bangladesh
[3]Department of Computer Science and Engineering, Dhaka University, Bangladesh

**Abstract:** Graphical passwords are implemented as an alternative scheme to replace alphanumeric passwords to help users to memorize their password. However, most of the graphical password systems are vulnerable to shoulder-surfing attack due to the usage of the visual interface. In this research, a method that uses shifting condition with digraph substitution rules is proposed to address shoulder-surfing attack problem. The proposed algorithm uses both password images and decoy images throughout the user authentication procedure to confuse adversaries from obtaining the password images via direct observation or watching from a recorded session. The pass-images generated by this suggested algorithm are random and can only be generated if the algorithm is fully understood. As a result, adversaries will have no clue to obtain the right password images to log in. A user study was undertaken to assess the proposed method's effectiveness to avoid shoulder-surfing attacks. The results of the user study indicate that the proposed approach can withstand shoulder-surfing attacks (both direct observation and video recording method). The proposed method was tested, and the results showed that it is able to resist shoulder-surfing and frequency of occurrence analysis attacks. Moreover, the experience gained in this research can be pervaded the gap on the realm of knowledge of the graphical password.

**Keywords:** graphical password, authentication, shoulder-surfing, digraph substitution rules, shifting condition

## 1. Introduction

Passwords are the most used means of identifying users in computer systems today. If the cost of equipment falls, biometric-based authentication (based on physiological activities such as fingerprints and facial recognition) will become more secure. Biometric solutions do not require memory and are quick and simple to implement (with cheap hardware). However, this strategy has significant disadvantages, such as the fact that users cannot be validated without the device, which can be stolen, lost, or rendered useless (e.g. when its battery runs out).

For the previous four decades, text-based password schemes have been a popular method of authentication for all users due to their ease, cost-effectiveness, simplicity, and familiarity (Shaikh et al., 2015). There are rules for creating a strong password, such as having at least eight characters, including a mix of alphabetic, numeric, and special characters, and using upper and lower case (Gayathiri, 2012). However, most users still use weak passwords to make it easier for them to remember them (Ho et al., 2014). A graphical password is introduced to help with memorability (Elftmann, 2006).

For authentication, graphical passwords are used instead of textual passwords. The limitation of password spaces is a well-known drawback of graphical password systems (Gao et al., 2013). Over the years, a large number of graphical password schemes have been developed. Most existing graphical password schemes, on the other hand, are vulnerable to a shoulder-surfing attack (Gao et al., 2010). Shoulder-surfing attacks are exacerbated by the visual interface and the limited amount of images used in graphical password systems (Biddle et al., 2012). As a result, the goal of this research is to offer a strategy that can prevent shoulder-surfing attacks while not compromising password strength when used in a system with restricted password spaces.

The contribution of this research is given below:

- To prevent shoulder-surfing attacks, a new method that uses shifting conditions with digraph substitution rules (DSR) was proposed.
- The proposed algorithm uses both password images and decoy images throughout the authentication procedure to confuse adversaries from obtaining the password images via direct observation or watching from a recorded session.

*Corresponding author: Amanul Islam, Department of Computer Science and Information Technology, University of Malaya, Malaysia. Email: aman.um16@siswa.um.edu.my

- Pass-images derived in this proposed algorithm are random and can only be derived with full knowledge of the algorithm.
- To be specific, adversaries will have no clue to obtain the right password images to login.
- Moreover, the proposed method is also immune to the frequency of occurrence analysis security threat even though a uniform randomization algorithm is used. The reason behind this is that the images used in all the challenge sets are similar.

Classification plays an important role in the field of graphical passwords. Recognition-based, recall-based, and cued recall-based graphical password schemes can be divided into three types (De-Angeli et al., 2005). During the password registration process, most recognition-based techniques ask users to memorize a portfolio of photos. Users must then identify the photos (password images/pass-images) before being able to log in. Users are needed to recall and replicate a secret drawing based on the registered images in recall-based methods. Users usually write their passwords on a blank canvas or a grid in these methods. On the other hand, cued recollection-based methods required users to recall and identify the target at precise positions within an image.

## 2. Related Work

Before the proposed method is presented in-depth, many selected recognition-based systems are identified and examined in terms of their strengths and drawbacks. The following is a review of some selected recognition-based systems.

Passfaces$^{TM}$ is a commercial product that was introduced by Passfaces$^{TM}$ Corporation in 2000 (Brosto and Sasse, 2000). It is the first recognition-based graphical password scheme. Users must choose human face photos as a password during the password registration step. Users must select the appropriate faces to log in throughout the login procedure. One of the benefits of the Passfaces$^{TM}$ method is that the password photos are simple to remember (Davis et al., 2004). This method, however, is susceptible to shoulder-surfing attacks.

In 2004, Davis et al. (2004) introduced the Story system. Instead of employing multiple categories of human face photos, the Story system uses a mixture of images from human and nonhuman face categories. Users must build their passwords by picking several photographs during the password registration step. Users must select their password images to log in during the authentication step. Users are urged to develop a tale to link their password images together to help them remember their passwords. Because Story uses the same authentication process as Passfaces$^{TM}$, it is susceptible to shoulder-surfing attacks.

In 2006, Weinshall created the Cognitive Authentication System (CAS). During the password registration step, users are requested to select multiple photos as their passwords (Weinshall, 2006). Users must concentrate on the top left image throughout the authentication step. Users must change their focused image to the bottom or one step below if the presently focused image is one of their password images. Otherwise, the next focused image will appear to the right of the present one. This procedure is repeated until the focused image is located at the grid's rightmost or bottommost point. The photos have numbers associated with them (ranging from zero to three), which are located at the rightmost and bottom of the grid. Users must identify the number that corresponds to the final focused image to log in. Users are not obliged to choose their password images or log in using a predetermined number connected with their password images under this system. As a result, their system can withstand shoulder-surfing attacks based on direct observation. It is not, however, impervious to record shoulder-surfing attacks, in which an attacker can watch numerous sessions and learn the user's password icons.

The Temporal Indirect Image-Based Authentication (TI-IBA) system was introduced by Yamamoto et al. (2009). Users must register various photos as their passwords during the password registration step. Users must identify their password images using a grid consisting of four slideshows throughout the authentication step. Every $t$ milliseconds, the images in the slideshows will change. Users must click on the slideshow containing their password photos to log in. The slideshows will play repeatedly to assist individuals who have forgotten their passwords. Because the TI-IBA system uses indirect password input to confound shoulder-surfing attackers, the authors claim that it can reduce direct observation shoulder-surfing attempts. The TI-IBA system is still vulnerable to shoulder-surfing assaults if the authentication process is captured.

In 2013, Por expanded the VIP3 system (Por, 2013). In this system, the author adjusted the frequency of particular distracter images so that they appeared more frequently than the others. The goal of such a suggestion is to persuade an opponent that the distracter image that appears more frequently is one of the password images (Por, 2013). The author employed partial selection and meta-heuristic procedures to choose the password and distracter images to prevent shoulder-surfing and FOA assaults. The suggested solution, according to the author, can prevent shoulder-surfing attacks and FOA because only a portion of the password images is revealed for each authentication challenge, and the appearance of the password images and distracter images is about the same on each challenge set. However, similar to the VIP3 system, the password can be discovered by analyzing numerous sessions.

Haque and Imam (2014) suggested a new graphical password in 2014, which combined recall and recognition-based techniques. Users must create a login and choose from a set of 25 photographs any number of images or a single image more than once during the registration step. The users are then free to choose any image from the stored image database or local memory. There will be a set of questions and graphics displayed. The users must choose three questions from the list. Each question must be paired with three photographs by the users. Users must identify the proper photos based on the question presented to log in. This strategy, according to the author, is simple to adopt and can avoid shoulder-surfing attacks. The password can still be discovered by observing numerous sessions.

Por et al. developed a graphical password method employing DSR to counter shoulder-surfing assaults in 2017 (Yee et al., 2017). Users must register a username and choose two photos as passwords throughout the registration process. To log in, the user must first learn and apply the three DSR. The technology, according to the authors, can withstand shoulder-surfing attacks. Furthermore, because all of the photos in the challenge set are identical, the author claims that the proposed method may avoid FOA attacks despite using a uniform randomization algorithm (Davis et al., 2004). However, DSR has a restriction in that the password images clicked by a user during a challenge set will lead to information about the password images used by the user. It means that the user's password images will be in the same column and row as the pass-images he or she has clicked. The password images can still be determined by filtering out the impossible images during several observations.

In 2019, Por et al. (2019) proposed LocPass, a new graphical password technique to prevent shoulder-surfing attacks. The user must create a User ID and validate it again as part of the registration process. The user receives a 5 × 5 grid after completing the User ID registration process. The user must register at least one place from the available grid. Additionally, the user is permitted to register more than once at the same place. The user must confirm the place they have chosen once they have made their choice. Once the registered locations are saved in the databases, the password registration process is deemed finished. The user must input the registered User ID during the authentication process. A 5 × 5 grid-based challenge set is then displayed after that. There are a total of 25 photos used. The images are chosen using a uniform randomness process, and the chosen images are then put into a 5 × 5 grid cell. The images used in LocPass have higher chances to offset each other. If an attacker tried to guess the recorded location utilized, offset may expand the password spaces of LocPass. However, if multiple sessions are observed, the password can still be found.

A shoulder-surfing-resistant scheme embedded in traditional textual passwords was proposed by Lai and Arko (2021). Users must build their passwords by picking several texts during the password registration step. In the authentication phase, A pattern appears in the password field as a suggestion to show a user how to enter a password when the password field is in focus. The user must type characters at the "O"s but not at the "X"s. There may be other characters in the password, but they were not part of the pattern, as shown by the "..." after the pattern. The user may quit after typing the entire password if it is shorter than the pattern. The proposed scheme asks users to skip 2–4 randomly selected characters in a password against shoulder-surfing attacks. Even if the input technique was captured, an observer cannot steal the full-length password because it is only entered partially. However, the password can still be discovered if numerous sessions are watched.

Khodadadi et al. (2021) proposed a novel graphical password authentication scheme with improved usability in 2021. There are three fields for gathering general information about users and their selected usernames during the registration phase. There is a 4 × 8 table which contains pictures of actors and football players who are well-known people. A single string with eight spaces is shown as a graphical password under the table of available alternatives. The drag and drop functionality allows users to select at least six images and add them to this string in any sequence they desire. The process of using a password is made simpler by the drag and drop capability. Users can pick between 6 and 8 images. Users must provide legitimate usernames and graphical passwords in order to log in successfully. So the system will redirect to the first page of the system, which can be whatever in terms of the usage of the system. The opacity of the user-selected images used as graphical passwords during login sessions will be lessened to help prevent some attacks, such as shoulder-surfing attempts. Nevertheless, the password may still be retrieved if more encounters are recorded.

When attackers watch several login sessions or the login sessions are recorded, most recognition-based systems are vulnerable to shoulder-surfing attacks, according to previous research. As a result, this study is being conducted to solve the shoulder-surfing attack problem in recognition-based systems, with a focus on multiple observations and video-recorded shoulder-surfing attacks. The proposed method is extensively explained in the following section.

## 3. Proposed Method

To test the proposed method, a prototype was created. The account setup procedure and the authentication procedure are both included in the prototype.

### 3.1. Procedure for account setup

**Figure 1**
**Process flow for the entire account setup procedure**

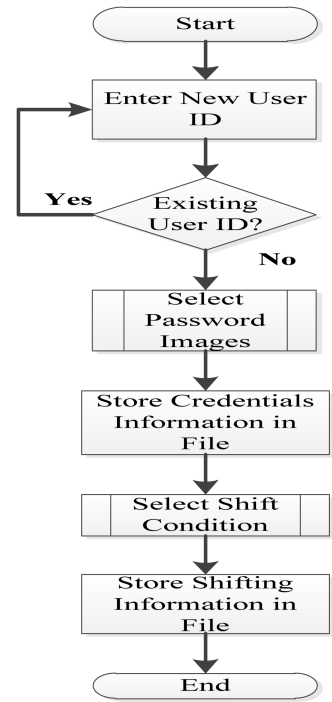

Figure 1 shows the process flow of the account setup procedure. There are three main steps for the account setup procedure which are given below:

- User ID creation
- Password images selection
- Shifting condition selection

Figure 1 shows the process flow for the entire account setup procedure. Initially, the user is required to create a new user ID using the proposed system. The system will check whether the user ID is occupied. The system will prompt an error message to the user if the user ID is occupied. After the user has successfully registered the user ID, the user is required to select his password. The user is required to select two images from 25 images in a 5 × 5 grid cell. An error message will be prompted to the user if the user has selected less than two or more than two images. Once the user has selected his password, the user ID and the password information will be stored in a system file. After that, the user is required to select a time unit and a shifting direction. Once the user has confirmed the shifting condition, the data of the selection options will be saved into a system file.

*3.1.1. User ID creation*

In the user ID creation step, a user is required to register the desired user ID. To ensure the user has keyed in the correct user ID, a "confirm user ID" field is used. The system will check whether the user ID is occupied in the storage file once the "OK" button is clicked. If the user ID has already existed, the system will prompt an error message to inform the user that the user ID was occupied and the user is required to register with another user ID.

**Figure 2**
**User interface for creating a new user ID**

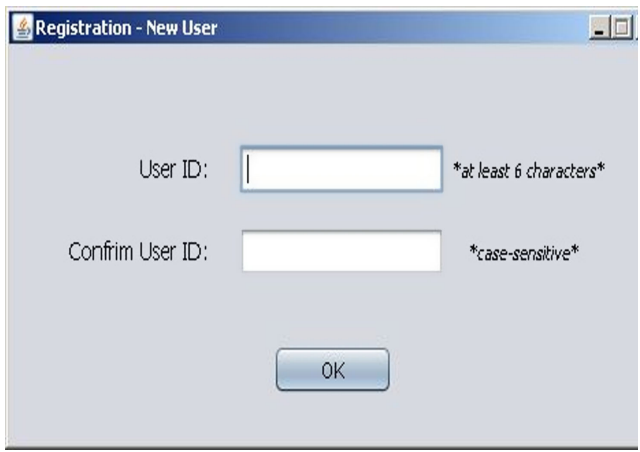

**Figure 3**
**Pseudocode for creating new user ID**

```
PROCEDURE UserID

    GET newUserID, newUserID2

    IF newUserID1 EQUALS newUserID2 THEN

        IF newUserID exists THEN

            Prompt user to provide another UserID

        ELSE

            Proceed to RegImage
```

**Figure 4**
**User interface for password image selection procedure**

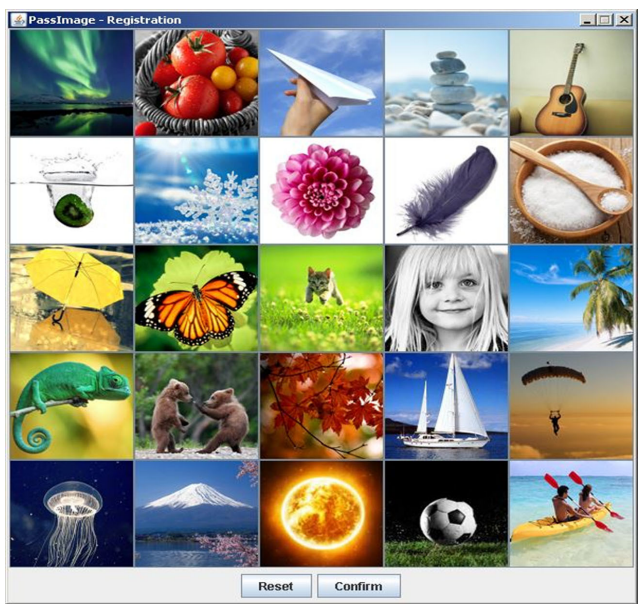

**Figure 5**
**Pseudocode for password image selection procedure**

```
PROCEDURE SelectPasswordImage

    SHOW ImageGrid

    IF ImageSelected NOT EQUALS 2 THEN

        SHOW WarningMessage

        GOTO SelectPasswordImage

    ELSE

        APPEND ImageSelected to File

        Back to MainMenu

    END IF

END PROCEDURE
```

The user can only proceed to the next step once the user has completed the user ID creation step. Figures 2 and 3 show the user interface and pseudocode to create a new user ID, respectively.

*3.1.2. Password images selection*

Following the completion of the user ID creation process, the user must choose a password. The login page consisted of 25 photographs in a 5 × 5 grid cell, and the user was forced to choose two images as his password (see Figure 4). All photographs are from the Internet stock source, and no copyright infringement is intended. If the user is unhappy with the selection, he or she can reset the password image choosing procedure. If the user is happy with the photographs they have chosen, they must click the "confirm" button to save them to the system file. If the user selects more than two or fewer two photos, an error warning will appear, and the user will be prompted to repeat the password choosing process. Figures 4 and 5 illustrate the user interface and the pseudocode for the password image selection procedure, respectively. Figure 6 illustrates the password images and their sequence for a predetermined user, which are labeled 1 and 2.

**Figure 6**
**A set of sample password images**

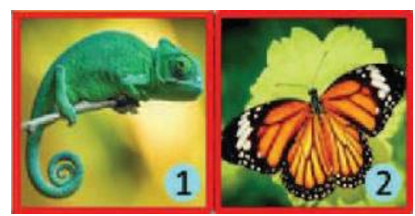

*3.1.3. Shifting condition selection*

After completing the password selection process, the user is brought to a new window where he has to select a shifting condition for his chosen password. In this step, the user is required to select one of the time units in the 24-h format time notation (HH: MM) and one of the shifting directions given (Up, Down, Left, and Right) (see Figures 7 and 8). The user can only select one of the time units and one of the shifting directions. The first hour time unit "$H_1$" ($H_1$ is denoted as the first hour time unit from $H_1H_2:M_1M_2$) and the "Up" direction of the shifting condition are selected for the user as a default setting. The system will save the selected parameters into a system file when the user clicks the "Confirm" button. Figures 7, 8 and 9 show the user interface and the pseudocode for the shifting condition selection, respectively.

**Figure 7**
**Shifting condition selection interface 1**

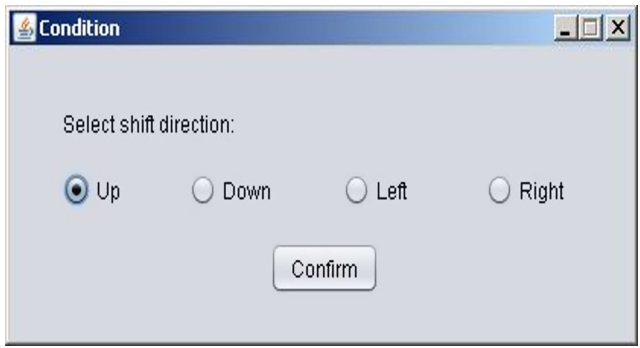

**Figure 8**
**Shifting condition selection interface 2**

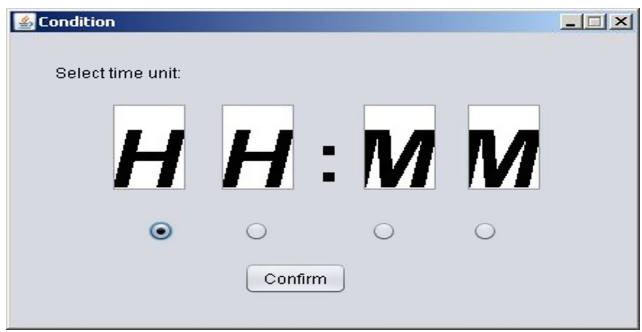

**Figure 9**
**Pseudocode for the shifting condition selection**

```
PROCEDURE SelectShiftCondition

        IF TimeUnit NOT selected OR ShiftDirection NOT
selected THEN

                SHOW WarningMessage

                GOTO SelectShiftCondition

        ELSE

                APPEND ShiftCondition to File

                Back to MainMenu

        END IF

END PROCEDURE
```

### 3.2. Authentication Procedure

**Figure 10**
**The authentication procedure's flow chart**

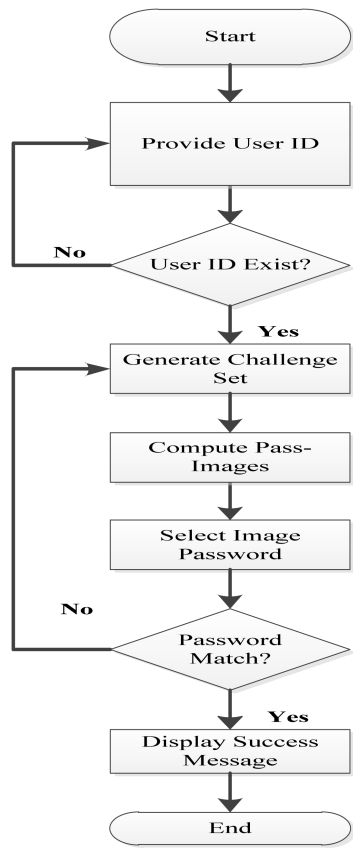

The authentication technique is depicted in Figure 10 as a flow chart. The authentication procedure is consisting of two steps:

- User ID verification
- Pass-images verification

*3.2.1. User ID verification*

The user must first enter the right User ID. An error message will be displayed if the user ID does not match the given User ID (see Figure 11). The user has to redo the user ID verification process. Otherwise, the user will proceed to the pass-images verification process.

**Figure 11**
**User interface for user ID verification**

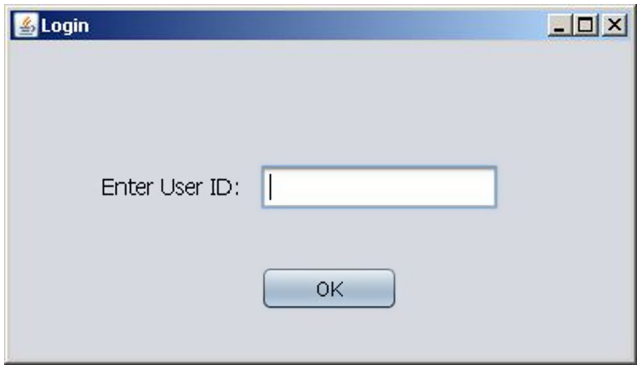

*3.2.2. Pass-images verification*

**Figure 12**
**User interface for pass-images verification**

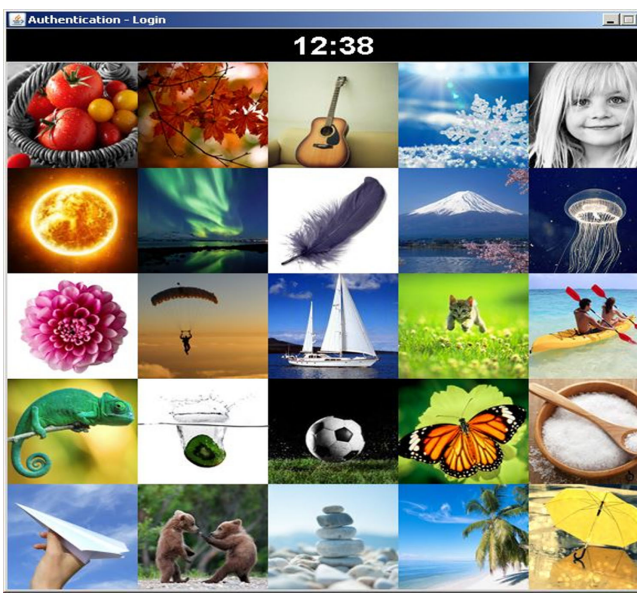

Following the user ID verification process, the user is immediately presented with a challenge set of 25 photos in a 5 × 5 grid cell (see Figure 12). Using the proposed technique, the user must identify the proper pass-images. The user can then log in to the prototype.

Three attempts are offered to the user to gain access to the prototype. After each failed effort, the user is given a fresh new challenge set. The new challenge uses the same photos, but the positions of each image will be distributed randomly using a uniform randomization technique. If the user fails to get authorized three times in a row, the user is denied access to the prototype.

If the user uses the suggested technique to correctly identify the pass-images, the prototype will display a successful login message, indicating that access has been given.

### 3.3. Proposed algorithm

The Playfair cipher, a manual symmetric encryption method, served as the primary inspiration for the suggested algorithm's fundamental concept. A user must comprehend three DSR in order to determine the pass-images.

A user must apply the three DSR proposed by Yee et al. (2017) along with the shifting condition to find the pass-images. DSR has three different possibilities.

Scenario A arises when both password images are diagonal to each other in general.

In Scenario A, the user must determine the row information and column information of the first pass-image using the first password image and the second password image, respectively. The first pass-image that the user must identify will be the intersection image. The same rule is used to create the second pass-image. The second password image is used to determine the row information for the second pass-image, whereas the first password image is used to determine the column information. The second pass-image that the user must identify is the intersection image.

Scenario B arises when both password images are on the same vertical vector. The first and second pass-images are right below the image in this case. There are several instances where the first or second password image is situated at the bottom of the column's side. The pass-image wraps over to the top side of the column in these exceptional circumstances.

Scenario C occurs when both password images are on the same horizontal vector. The first and second pass-images in this circumstance are their immediate right pictures, respectively. The pass-image wraps around the left side of the row if the first or second password image is positioned at the right edge of the row. Yee et al. (2017) provides a detailed coordinate presentation for determining the location of the pass-images in the three circumstances.

Additionally, to strengthen the rule, an enhancement feature is added to the core algorithm. A numerical shift has been introduced to the DSR algorithm. With this feature, the user is required to select a time unit from a 24-h clock format and one of the shifting directions. Finally, the user will have to decide the pass-images using these shifting circumstances and the digraph replacement criteria during the authentication method. The following equations can be used to determine the final placement of the pass-images:

Direction: Up

$$P1_{xy} = (P1_x,\ P1_{(y+T)\ mod\ my}) \quad (1)$$

$$P2_{xy} = (P2_x,\ P2_{(y+T)\ mod\ my}) \quad (2)$$

Direction: Down

$$P1_{xy} = (P1_x,\ P1_{(y-T)\ mod\ my}) \quad (3)$$

$$P2_{xy} = (P2_x,\ P2_{(y-T)\ mod\ my}) \quad (4)$$

*my* is the maximum range of the y-axis of a (m x m) or a (m x n) grid cell, *T* is the value of the time unit selected by the user during the account setup procedure, and *mod* is a modulo of a number.

Direction: Right

$$P1_{xy} = (P1_{(x+T)\ mod\ mx},\ P1_y) \quad (5)$$

$$P2_{xy} = \left(P2_{(x+T) \bmod mx},\ P2_y\right) \quad (6)$$

Direction: Left

$$P1_{xy} = \left(P1_{(x-T) \bmod mx},\ P1_y\right) \quad (7)$$

$$P2_{xy} = \left(P2_{(x-T) \bmod mx},\ P2_y\right) \quad (8)$$

*mx* is the maximum range of the x-axis of a (m x m) or a (m x n) grid cell, *T* is the value of the time unit selected by the user during the account setup procedure, and *mod* is a modulo of a number.

Assume a user has saved the password images and the shifting condition as shown in Figures 6, 7 and 8, respectively. The sequence of the password images is marked as 1 and 2. The registered shifting direct is "Up" and the time unit selected is the first hour time unit, "$H_1$". Figure 13 depicts an example of a single challenge round. The intermediate pass-images are indicated in yellow boxes after being identified using the DSR, and their sequences are labeled P1 and P2. The pass-images are outlined in green boxes, and their sequences are labeled N1 and N2.

The row and column information for identifying the pass-images is indicated by the blue and green arrows.

Before using any rule to find the pass-images, the user must first discover the location of the password images in a challenge set. Both password images appear to be horizontal to each other in the figure (Scenario C). P1 and P2 can be calculated by identifying the immediate right images next to the first and second password images, respectively, or by using the following equations (Yee et al., 2017):

$$P1_{xy} = \left(P1_{(x+1) \bmod mx},\ P2_y\right) \quad (9)$$

$$P2_{xy} = \left(P1_{(x+1) \bmod mx},\ P2_y\right) \quad (10)$$

From Figure 13, the first hour is the number "1" in "12:38". Thus, the first final pass-image, N1, is one image on top of P1 and the second final pass-image, N2, is also one image on top of P2 (or use equations (1) and (2)). The user must click on the pass-images to finish the challenge round after identifying the final pass-images. Algorithm 1 displays the suggested algorithm's pseudocode.

**Figure 13**
**An example of a challenging round**

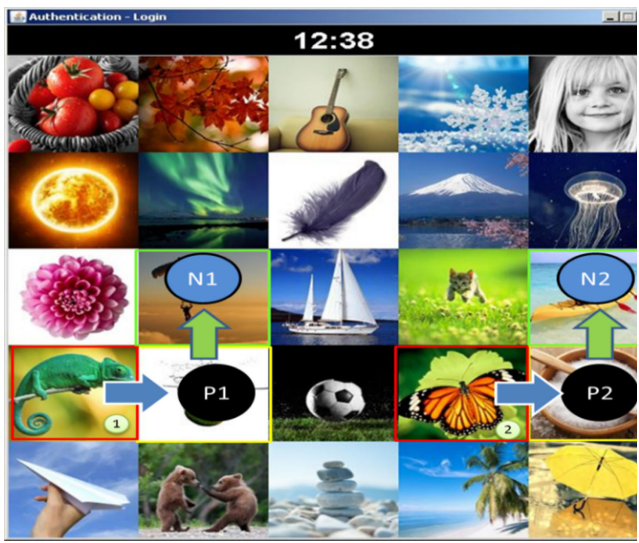

**Algorithm 1** Pseudocode for the proposed algorithm

1: OPEN file // Open file containing a list of user's id and password images
2: SET loginID = user's input of string during login
3: SET v = new empty vector // New vector to store user's password image descriptions from file
4: CHECK file // Scans the file
5: IF loginID is found THEN
6: Launch authentication window // User ID authentication success
7: SCANS next line for password image
8: ADD v = user's pass-image
9: ELSE
10: Launch warning window // User ID does not exist
11: END IF
12: SET coorX = new integer list for x coordinates
13: SET coorY = new integer list for y coordinates
14: SET m = 0
15: SET m < 2
16: FOR the number of images on the grid
17: GET description of each images
18: IF description of image = v[m]
19: ADD coorX = image's x coordinate
20: ADD coorY = image's y coordinate
21: END IF
22: END FOR
23: INCREMENT m
24: END FOR
25: SET X1 = x coordinate of 1$^{st}$ password image
26: SET X2 = x coordinate of 2$^{nd}$ password image
27: SET Y1 = y coordinate of 1$^{st}$ password image
28: SET Y2 = y coordinate of 2$^{nd}$ password image
29: IF Y1>Y2 or Y2>Y1 THEN // Scenario A
30: SET passX1 = X2
31: SET passY1 = Y1
32: SET passX2 = X1
33: SET passY2 = Y2
34: ELSE IF Y1=Y2 THEN // Scenario C
35: SET passX1 = X1 + 1
36: SET passX2 = X2 + 1
37: IF X1 > 4 THEN // Special case
38: SET passX1 = my − X1
39: ELSE IF X2 > 4 THEN
40: SET passX2 = my − X2
41: END IF
42: ELSE IF X1=X2 THEN // Scenario B
43: SET passY1 = Y1 + 1
44: SET passY2 = Y2 + 1
45: IF Y1 > 4 THEN // Special case
46: SET passY1 = mx − Y1
47: ELSE IF Y2 > 4 THEN
48: SET passY2 = mx − Y2
49: END IF
50: END IF
51: condDirection = {UP, DOWN, LEFT, RIGHT}
52: condTimeUnit ={H1, H2, M1, M2}
53: condSelected ={DirectionSelected, TimeUnitSelected}
54: T= TimeUnitSelected
55: IF DirectionSelected = UP THEN
56: npassX1=passX1
57: npassY1=(passY+T) mod max Y
58: npassX2=passX2
59: npassY2=(passY+T) mod max Y

(*Continued*)

**(***Continued* **)**

| **Algorithm 1** Pseudocode for the proposed algorithm |
|---|
| 60: IF DirectionSelected = DOWN THEN |
| 61: npassX1=passX1 |
| 62: npassY1=(passY-T) mod max Y |
| 63: npassX2=passX2 |
| 64: npassY2=(passY-T) mod max Y |
| 65: IF DirectionSelected = UP THEN |
| 66: npassX1=passX1 |
| 67: npassY1=(passY+T) mod max Y |
| 68: npassX2=passX2 |
| 69: npassY2=(passY+T) mod max Y |
| 70: IF DirectionSelected = DOWN THEN |
| 71: npassX1=passX1 |
| 72: npassY1=(passY-T) mod max Y |
| 73: npassX2=passX2 |
| 74: npassY2=(passY-T) mod max Y |
| 75: END IF |

## 4. User Study

### 4.1. Participants

User research was undertaken to see if the proposed strategy was effective in minimizing shoulder-surfing attacks. This user study enlisted the participation of 140 students from the University of Malaya's Faculty of Computer Science and Information Technology; 90 were male and 50 were female. The participants in the user study were not compensated for their time.

### 4.2. Procedure

The user research required participants to devote a maximum of 15 min to it. They were first given the notion of the proposed method as well as the procedures they need take to log in. Before creating their password and logging into the prototype, the participants were instructed to go through the instruction. Participants were requested to choose two photographs as their password images during the registration phase. To familiarize themselves with the proposed strategy, they were instructed to utilize the password images for 10 successful login attempts. The prototype kept track of how long it took to log in. After completing 10 successful logins, the participants were instructed to watch a login session demonstration and execute a shoulder-surfing test. The photos provided in the challenge set and the images clicked by the authorized user upon login are the resources available to them as attackers. Then they were given an unlimited number of chances to guess the password photos used in the demonstration login session. They were then given a posttest questionnaire to obtain comments on their shoulder-surfing technique. Each person was tested individually.

## 5. Experimental Results

Participants were asked a few questions before the login training and demonstration to determine their general grasp of graphical passwords and shoulder-surfing assaults. They'd all heard of a graphical password, and everyone had heard of a shoulder-surfing attack. About 86 percent of the participants had prior security hacking experience, while the other 14 percent did not.

Figure 14 depicts 10 successful logins in the interim. Over the 10 login attempts, the time it took the participants to log in fell dramatically, as shown in the graph. This demonstrated that as participants gained more comfortable with the prototype, the time required to log in decreased.

**Figure 14**
**The average time for 10 successful logins in seconds**

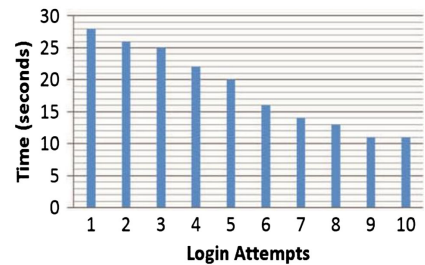

**Table 1**
**Statistics of successful login time**

| Item | Time (s) |
|---|---|
| Minimum | 7.0 |
| Maximum | 32.0 |
| Average | 18.6 |
| Standard deviation | 6.1 |
| Median | 17.0 |

For all successful login attempts, Table 1 provides the lowest, maximum, average, standard deviation, and median. The participants took an average of 7.0 s to successfully log in, while the greatest duration was 32.0 s. The participants took an average of 18.6 s to successfully log in, with a standard deviation of 6.1 s. The average time it takes to log in successfully is 17.0 s. This indicates that 50% of the login attempts took less than 17.0 s to complete.

After completing 10 successful logins, the participants were instructed to watch a login session demonstration. Participants were free to see the login session demonstration as many times as they liked. Then they were given an infinite number of trials to figure out the password photos used in the sample login session. None of the participants in the shoulder-surfing test were able to recognize the password images used in the demonstration session. One would think that knowing the algorithm utilized would offer the opponent an advantage in carrying out their assaults.

Even though the attackers were aware of the proposed system and the underlying algorithm, the findings of the shoulder-surfing test in the user study revealed that the proposed technique is resistant to shoulder-surfing attacks (both traditional and video-recorded shoulder-surfing attacks).

The participants were given a posttest questionnaire in which they were asked about their password-gathering approach. Figure 15 depicts the participants' techniques and the percentage of individuals who employed them. Despite knowing how the underlying proposed algorithm works, 86.8% of the participants were unable to figure out the pass-images employed. 9.4% of the participants just guessed the pass-pictures, whereas 3.8% used direct observation and clicked following the images given in the demonstration.

**Figure 15**
**Participants' strategies for obtaining pass-images**

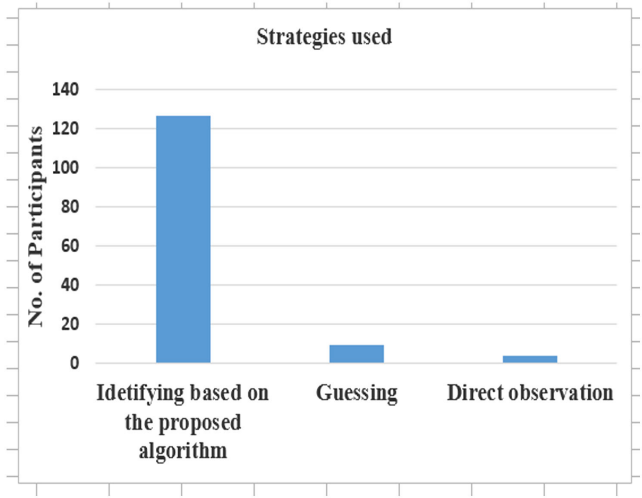

**Figure 16**
**User memorability test result**

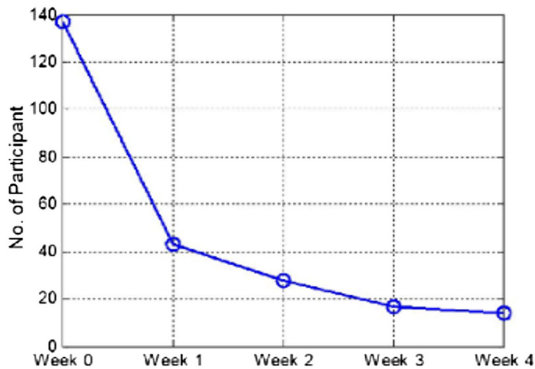

**Table 2**
**Comparison of the related work**

| Graphical password Scheme | Resist DO | Resist MOs | Resist VRSSA |
|---|---|---|---|
| Passfaces[TM] (Brosto and Sasse, 2000) | NO | NO | NO |
| Story (Davis et al., 2004) | NO | NO | NO |
| CAS (Weinshall, 2006) | YES | NO | NO |
| TI-IBA (Yamamoto et al., 2009) | YES | NO | NO |
| Por (2013) | YES | NO | NO |
| Haque and Imam (2014) | YES | NO | NO |
| DSR (Yee et al., 2017) | YES | NO | NO |
| LocPass (Por et al., 2019) | YES | NO | NO |
| Lai and Arko (2021) | YES | NO | NO |
| Khodadadi et al. (2021) | YES | NO | NO |
| Proposed method | YES | YES | YES |

DO: direct observation, MOs: multiple observations, VRSSA: video-recorded shoulder-surfing attack.

Figure 16 shows the user memorability test result of our proposed method. About 31% of the participants can still remember their registered images and use our proposed method to login a week after they have completed the shoulder-surfing test. In week 2, the percentage drops to 20% followed by 12% and 10% in week 3 and week 4 respectively. According to the participants, the main reason that they failed to login is because they cannot remember the registered images and also they have forgotten how our proposed method works. From this test, we found out that although our proposed method can resist shoulder-surfing attack, but there is a trade-off in terms of user memorability. In future, we might need to revise the images used and also our proposed method so that it can assist the user to recall their password better.

## 6. Discussion

The comparison of password systems is shown in Table 2. The proposed strategy is the only system that can avoid shoulder-surfing attacks, according to the table. The authors employed the notion of the indirect input method to prevent direct observation shoulder-surfing attacks in the CAS (Weinshall, 2006) and TI-IBA (Yamamoto et al., 2009) systems. To prevent direct observation shoulder-surfing attacks, the inventors of Por (2013) and Haque and Imam (2014) systems adopted the notion of partial password selection method. When several login sessions are logged and monitored, the password images can still be revealed. Attackers can get the password images by removing the fake images from each challenge set. The password images clicked by the user during a challenge set will lead to information about the password images used by the user, according to DSR (Yee et al., 2017). Based on the three situations (Scenario A, B, and C) utilized in DSR, the user's password images will fall within the same column and row of the pass-pictures clicked by the user. The password images can still be determined by filtering out the impossible images during several observations.

A uniform randomization mechanism is employed in the suggested method to show the same images in each challenge set. As a result, the suggested system is immune to FOA attacks. In terms of shoulder-surfing resistance, the proposed strategy employs the DSR in conjunction with the shifting condition to deceive adversaries into detecting the right pass-images. Because the shifting condition in each task set varies depending on the direction and time unit chosen. Even though the attackers knew about the DSR and which scenario information is utilized in each challenge set, shoulder-surfing attackers were unable to get knowledge about the password images and pass-images the user used.

Even though the shoulder-surfing attack was defeated, there is still potential for improvement. As noted in Yee et al. (2017), research can be conducted on overcoming various security threats. If a graphical password is to be used in an online system, research into overcoming social engineering, phishing, pharming, and man-in-the-middle attacks should be explored. The methods of data concealing and sophisticated cryptography presented in Por and Delina (2008), Izadeen and Ameen (2021), Yang (2019), Meng et al. (2017) could be used to address the aforementioned issues. Our main focus will be on the use of falsified movement to deceive attackers, as indicated in Meng et al. (2019), as well as the development of other security considerations such as password length increments and usability in terms of assisting in password memorability.

## 7. Conclusion

A new shoulder-surfing resistant strategy based on changing conditions and DSR is proposed in this work. The proposed method was put to the test, and the findings demonstrate that it can withstand attacks from direct observation, multiple observation, and video recording. Users were able to type their passwords accurately and remember them over time, according to usability testing of the suggested technique. Fifty percent of participants took 17.0 s to log in during the first 10 successful logins.

Furthermore, the knowledge collected in this study can be used to bridge the knowledge gap in the field of graphical passwords.

## Conflicts of Interest

The authors declare that they have no conflicts of interest to this work.